\documentclass[dvips]{article}
\usepackage{graphicx}
\newcommand{\doublespacing}{\let\CS=\@currsize\renewcommand{\baselinesstrech}
{2.0}\tiny\CS}

\begin{document}

\textwidth 16cm
\newcommand{\bd}{\begin{document}}
\newcommand{\ed}{\end{document}}
\newcommand{\bc}{\begin{center}}
\newcommand{\ec}{\end{center}}
\newcommand{\bfr}{\begin{flushright}}
\newcommand{\efr}{\end{flushright}}
\newcommand{\lt}{\left}
\newcommand{\rt}{\right}
\newcommand{\vs}{\vspace}
\newcommand{\hs}{\hspace}
\newcommand{\beq}{\begin{equation}}
\newcommand{\eeq}{\end{equation}}
\newcommand{\lb}{\linebreak}
\newcommand{\pb}{\pagebreak}
\newcommand{\mb}{\makebox}
\newcommand{\fb}{\framebox}
\newcommand{\mc}{\multicolumn}
\newcommand{\ben}{\begin{enumerate}}
\newcommand{\een}{\end{enumerate}}
\newcommand{\bit}{\begin{itemize}}
\newcommand{\eit}{\end{itemize}}
\newcommand{\ol}{\overline}
\newcommand{\un}{\underline}
\newcommand{\lefq}{\lefteqn}
\newcommand{\ba}{\begin{array}}
\newcommand{\ea}{\end{array}}
\newcommand{\beqa}{\begin{eqnarray}}
\newcommand{\eeqa}{\end{eqnarray}}
\newcommand{\beqas}{\begin{eqnarray*}}
\newcommand{\eeqas}{\end{eqnarray*}}
\newcommand{\bfg}{\begin{figure}}
\newcommand{\efg}{\end{figure}}
\newcommand{\bds}{\begin{displaymath}}
\newcommand{\eds}{\end{displaymath}}
\newcommand{\btb}{\begin{tabbing}}
\newcommand{\etb}{\end{tabbing}}
\newcommand{\para}{\parallel}
\newcommand{\pad}{\partial}
\newcommand{\nn}{\nonumber}
\newcommand{\la}{\leftarrow}
\newcommand{\ra}{\rightarrow}
\newcommand{\lgla}{\longleftarrow}
\newcommand{\lgra}{\longrightarrow}
\newcommand{\La}{\Leftarrow}\newcommand{\Ra}{\Rightarrow}
\newcommand{\Lra}{\Leftrightarrow}
\newcommand{\Lgla}{\Longleftarrow}
\newcommand{\Lgra}{\Longrightarrow}
\newcommand{\bm}{\boldmath}
\newcommand{\lan}{\langle}
\newcommand{\ran}{\rangle}
\renewcommand{\a}{\alpha}
\renewcommand{\b}{\beta}
\newcommand{\g}{\gamma}
\newcommand{\G}{\Gamma}
\renewcommand{\d}{\delta}
\newcommand{\eps}{\epsilon}
\newcommand{\Th}{\Theta}
\newcommand{\s}{\sigma}
\newcommand{\lam}{\lambda}
\newcommand{\D}{\Delta}
\newcommand{\vare}{\varepsilon}
\newcommand{\pr}{\prime}
\newcommand{\ro}{\rho}
\newcommand{\nab}{\nabla}
\newcommand{\m}{\mu}
\newcommand{\n}{\nu}
\newcommand{\Sg}{\Sigma}
\newcommand{\p}{\pi}
\newcommand{\R}{I\!\!R}
\newcommand{\om}{\omega}
\newcommand{\Om}{\Omega}
\newcommand{\ze}{\zeta}
\newcommand{\vart}{\vartheta}
\newcommand{\tri}{\triangle}
\newcommand{\f}{\frac}
\newcommand{\iny}{\infty}
\newcommand{\pro}{\propto}
%\input{fqhsc}
%\input{state}
%\input{acc}
%\input{fs}
%\ed
\bc {\huge \bf $\cal{PT}$ symmetric models with nonlinear pseudo supersymmetry} \ec

\vs{1cm}

\bc
{\it A. Sinha{\footnote {e-mail : anjana23@rediffmail.com}}\\
Department of Applied Mathematics \\
Calcutta University \\
92, APC Road, Kolkata - 700 009, India.}
\ec

\vs{.25cm}

\bc
{and}
\ec

\vs{.25cm}

\bc
{\it P. Roy{\footnote {e-mail : pinaki@isical.ac.in}} \\
Physics \& Applied Mathematics Unit \\
Indian Statistical Institute \\
Kolkata - 700 108, India.}
\ec

\vs{1cm}

\vs{1cm}

\bc
{\large {\un{Abstract}}}
\ec

By applying the higher order Darboux algorithm to an exactly
solvable  non Hermitian ${\cal{PT}}$ symmetric potential, we
obtain a hierarchy of new exactly solvable non Hermitian
${\cal{PT}}$ symmetric potentials with real spectra. It is shown
that the symmetry underlying the potentials so generated and the
original one is {\it nonlinear pseudo supersymmetry}. We also show
that this formalism can be used to  generate a larger
 class of new solvable potentials when applied to non Hermitian systems. 

\pb

\section{Introduction}
There are not many exactly solvable potentials in quantum
mechanics. As a result there have always been efforts to enlarge
the class of exactly solvable potentials. Some of the different
methods which have been used time and again to generate a
hierarchy of isospectral potentials are the factorization method
of Infeld and Hull \cite{hull}, the Darboux algorithm \cite{dar},
the method of supersymmetric quantum mechanics (SUSY QM)
\cite{susy}, or the integral transformations of
Abraham-Moses-Pursey \cite{am} etc. Among these methods the
Darboux algorithm and the SUSYQM are closely related and these
methods have found numerous applications in different areas of
theoretical and mathematical physics \cite{susy}.

At the same time, the scheme is still narrow as conventional SUSY
fails to explain certain phenomena, e.g., the disappearance of the
leading Borel singularity of the perturbation correction for the
ground state energy of a SUSY theory \cite{aoyama}. In order to
explain such behaviour and also to widen the scope of SUSY QM, an
idea was put forward to extend SUSY to higher orders
\cite{andrianov}. We recall that in the conventional intertwining
technique, two one-dimensional Schr\"{o}dinger Hamiltonians $H$
and $ {\widetilde{H}}$ are intertwined by means of differential
operators $L$ as \beq {\widetilde{H}} L = L H \ \ \ \ \ \ \ \ \ \
\ \ \ H L^{\dag} = L^{\dag} {\widetilde{H}}  \eeq If $L$ is of the
first order in derivatives, the standard SUSY QM, with
supercharges built of first order Darboux transformation
operators, and the factorization method are recovered. On the
other hand, if higher order differential operators are involved in
the construction of $L$, it is variously referred to as {\it
polynomial SUSY} \cite{andrianov}, or {\it nonlinear SUSY}
\cite{plyush}, or {\it higher order SUSY ($n$-SUSY)}
\cite{fernan,samsonov},  or ${\cal{N}}${\it fold SUSY}
\cite{aoyama,tanaka}, the study of which has attracted the
attention of a lot of researchers in recent times
\cite{aoyama,andrianov,plyush,fernan,samsonov,tanaka}. Contrary to
standard SUSY, the anticommutator of the supercharges no longer
coincides with the Hamiltonian in general. Instead, it becomes a
polynomial of the Hamiltonian in degree $N$, and is sometimes
referred to as the {\it Mother Hamiltonian} \cite{aoyama,tanaka}.

Furthermore, the equivalence between an $N$-th
order Darboux transformation and a chain of $N$
first order Darboux transformation is well
established \cite{samsonov}. Every chain of $N$
first order Darboux transformation creates a
chain of exactly solvable Hamiltonians $ h_0
\rightarrow h_1 \rightarrow \cdots \rightarrow
h_N $. Hence the intertwining operator $L^{(N)}$
between the initial Hamiltonian $h_0$ and the
final Hamiltonian $h_N$ can always be presented
as a product of $N$ first order Darboux
transformation operators between every two
juxtaposed Hamiltonians $h_0, h_1, \ldots, h_N$ :
\beq L^{(N)} = L_N L_{N-1} \cdots L_2 L_1 \qquad
h_p L_p = L_p h_{p-1} \qquad p = 1,2, \cdots N
\eeq In conventional higher order SUSY, $h_0$ and
$h_N$ are essentially self-adjoint Hermitian
operators in a Hilbert space, with square
integrable eigenfunctions. If all the
intermediate potentials $V_1 (x), V_2 (x),
\ldots, V_{N-1} (x) $ are real valued functions
in their common domain of definition $(a,b)$, the
chain is called {\it reducible}, and the $N$-th
order Darboux transformation is called {\it
reducible} as well. Additionally, if all the
intermediate potentials are free of singularities
in $(a,b)$, the chain and the corresponding
transformation are called {\it completely
reducible}. When at least one intermediate
potential is a complex valued function, the chain
and the corresponding transformation are called
{\it irreducible}.

At the same time, non Hermitian Hamiltonians have
made an important place for themselves in the
recent development of quantum mechanics, because
of their intrinsic interest \cite{pt} and
possible applications \cite{hatano}. It is well
known by now that a non Hermitian ${\cal{PT}}$
symmetric Hamiltonian admits real eigenvalues if
the eigenfunctions, too, respect the ${\cal{PT}}$
invariance (the so-called unbroken ${\cal{PT}}$
symmetry), whereas the eigenvalues occur as
complex conjugate pairs if ${\cal{PT}}$ symmetry
is spontaneously broken (in this case the
eigenfunctions are no longer ${\cal{PT}}$
invariant). For such non Hermitian ${\cal{PT}}$
symmetric Hamiltonians, \beq {\cal{PT}} H = H
{\cal{PT}} \label{ptsym} \eeq where ${\cal{P}}$
stands for the {\it space inversion} operator and
${\cal{T}}$ denotes {\it time reversal} : \beq
\lt. \ba {lcl}
{\cal{P}} ~ &:& ~ x \rightarrow -x, \qquad p \rightarrow -p  \\
{\cal{T}} ~ &:& ~ x \rightarrow x, \qquad p \rightarrow -p, \qquad
i \rightarrow -i  \ea \rt\} \label{pt} \eeq The reality of the
spectrum may be attributed to the so-called $\eta$-pseudo
Hermiticity of the non Hermitian Hamiltonian \cite{mostafa} \beq
H^{\dagger} = \eta H \eta ^{-1} \eeq where $ \eta $ is a linear,
invertible, Hermitian operator. Several non-Hermitian
Hamiltonians, whether possessing ${\cal{PT}}$ invariance or not,
have been identified as $\eta$ pseudo-Hermitian under $ \eta =
e^{- \theta p}$ , where $ \theta $ is real, and $ p ~=~ -i
\f{d}{dx} $, or $ \eta = e^{- \phi (x) }$, where $ \phi (x) $ is
some gauge-like transformation. We note that for ${\cal{PT}}$
symmetric Hamiltonians, $ \eta $ may simply taken as the parity
operator ${\cal{P}}$, whereas for conventional Hermitian
Hamiltonians, $ \eta ~=~ 1$.

Moreover, the square integrability of the wave functions is no
longer a pre-requisite for non Hermitian Hamiltonians. Instead,
the ortho-normalization of the wave function for Hermitian quantum
mechanics \beq \int \Psi _m ^* ~ \Psi _n  dx = \delta _{m,n} \eeq
is replaced by \cite{c-op} \beq \int \lt[ {\cal{CPT}} \Psi _m \rt]
\Psi _n dx = \delta _{m,n} \label{cpt} \eeq where ${\cal{C}}$
plays the role of a linear charge operator, obeying the
relationship \beq \lt[ {\cal{C}},H \rt] = 0 \ \ \ \ \ \ \ \lt [
{\cal{C,PT}} \rt] = 0 \eeq and has the property $ {\cal{C}} ^2 = 1
$. In the position representation ${\cal{C}}$ is given as \beq
{\cal{C}} (x,y) = \sum _n \psi _n (x) \psi _n (y) \eeq and the
completeness relation gets modified to \beq \sum _n \lt [
{\cal{CPT}} \psi _n (x) \rt] \psi _n (y) = \delta (x-y)  \eeq

While nonlinear SUSY for $N=2$, has been investigated widely for
Hermitian Hamiltonians [5-10], such studies have not been carried
out as yet for non Hermitian Hamiltonians. Motivated by the
importance of such systems in the recent development of quantum
mechanics, our aim in the present work is to generalise the
concept of nonlinear SUSY to include non Hermitian quantum
systems. In analogy with the first order systems, where the
partner Hamiltonians $H_{\pm}$ of non Hermitian systems were found
to be related through {\it pseudo supersymmetry}
\cite{mostafa,jpa}, it will be shown that the underlying symmetry
between the isospectral partners $h_0$ and $h_N$ is a
generalisation of ${\cal{N}}$ SUSY and may be called {\it
nonlinear  pseudo  supersymmetry}. The nature of the intermediate
Hamiltonians as well as the corresponding wave functions will also
be investigated.

The organization of the paper is as follows. For the sake of
completeness, in section $2$ we briefly outline conventional
nonlinear SUSY for Hermitian quantum mechanics. In section $3$ we describe a similar framework
for non-Hermitian Hamiltonians and show that,
the underlying symmetry for the potentials produced by higher order Darboux algorithm,
is nonlinear pseudo supersymmetry. Some
explicit examples are given in sections $4$ and $5$ , while
section $6$ is devoted to a conclusion.

\section{Non linear SUSY for Hermitian Hamiltonians}

In the conventional first order supersymmetric quantum mechanics,
if a given solvable Hamiltonian \beq H = - \frac{d^2}{dx ^2} +
V(x) \eeq possesses a discrete spectrum of bound states $E_n, \ \
n=0,1,2, \cdots $, together with the square-integrable
eigenfunctions $ \psi _n (x) $, then a pair of first-order
operators $L_0$ and $L_0 ^{\dagger}$ can be constructed from the
ground state $\psi _0$, given by \beq L_0 = \frac{d}{dx} + W_0
(x), \ \ \ \ \ \ L_0 ^{\dagger} = - \frac{d}{dx} + W_0 (x) \eeq
where \beq W_0 (x) = - \left[ \ln \psi _0 (x) \right] ^{\prime}
\eeq such that $L_0$ and $L_0 ^{\dagger} $ play the role of
intertwining operators for the initial and final Hamiltonians $H$
and $ \widetilde{H} $, respectively : \beq {\widetilde{H}} L_0 =
L_0 H \ \ \ \ \ \ \ \ \ \ \ \ \  H L_0 ^{\dag} = L_0 ^{\dag}
{\widetilde{H}} \label{intertwine} \eeq with \beq H = L_0 ^{\dag}
L_0, \ \ \ \ \ \ \ \widetilde{H} = L_0 L_0 ^{\dag} \eeq Simple
straightforward algebra shows that the partner potentials $V(x)$
and $\widetilde{V}(x)$ can be expressed as \beq V(x) = W_0 ^2 (x)
- W_0 ^{\prime} (x) \eeq \beq \widetilde{V}(x) = W_0 ^2 (x) + W_0
^{\prime} (x) = V(x) + 2 W_0 ^{\prime} (x) \eeq The eigenfunctions
$ \psi (x) $ and $ \widetilde{ \psi} (x) $ of $H$ and
$\widetilde{H}$ are interrelated through $L_0$ and $L_0 ^{\dagger}
$ : \beq L_0 \psi _0 (x) = 0, \ \ \ \ \ \ L_0 (x) \psi _i (x) =
\frac{W_{0,i} (x)}{\psi_0 (x) }  ~~ \a  ~~  \widetilde{\psi} _i
(x), \ \ \ \ \ L_0 ^{\dagger} \widetilde{\psi} _i (x) ~~  \a ~~
\psi _i (x) , \ \ \ \ \ i = 1,2,\cdots  \eeq where $W_{0,i} (x) =
\lt\{ \psi _0 (x) \psi _i ^{\prime} (x) - \psi _0 ^{\prime} (x)
\psi _i (x) \rt\} $ is the Wronskian of $ \psi _0 (x) $ and $
\psi_i (x) $. The concise algebraic form of spectral equivalence
is given by the superalgebra for the partners $H$ and
$\widetilde{H}$, and the
supercharges $Q$ and $Q^{\dagger}$ : \beq Q = \left(%
\begin{array}{cc}
  0 & L_0 \\
  0 & 0 \\
\end{array}%
\right), \ \ \ \ \ \ \ \ \ \ \ \ \ \ Q ^{\dagger} = \left(%
\begin{array}{cc}
  0 & 0 \\
  L_0 ^ {\dagger} & 0 \\
\end{array}%
\right) \eeq
\beq {\cal{H}} = \lt\{ Q, Q^{\dagger} \rt\} = \left(%
\begin{array}{cc}
  H & 0 \\
  0 & \widetilde{H}  \\
\end{array}%
\right)
= \left(%
\begin{array}{cc}
  L_0 ^{\dagger} L_0 & 0 \\
  0 & L_0  L_0 ^{\dagger}  \\
\end{array}%
\right) \eeq satisfying the relations \beq \lt\{ Q, Q \rt\} =
\lt\{ Q^{\dagger}, Q ^{\dagger} \rt\} =0, \ \ \ \ \    \lt[ Q,
{\cal{H}} \rt] = \lt[ Q ^{\dagger} , {\cal{H}} \rt] = 0 \eeq Thus
$H$ and $ \widetilde{H}$ are isospectral except for the lowest
eigenvalue $E_0$ which is missing in $\widetilde{H}$, as $
\widetilde{ \psi}_0 $ is not normalizable.

To generalize standard SUSY to higher order, the
supercharges are built of higher order
intertwining operators \cite{samsonov}. The two
Hamiltonians $h_0$ and $h_N$ are intertwined
through an $N$-th order differential operator
$L^{(N)}$, as \beq L^{(N)} h_0 = h_N L^{(N)}, \ \
\ \ \ \ \ h_0 L^{(N) \dagger} = L^{(N) \dagger}
h_N \eeq where $h_0$ and $h_N$ are self adjoint
operators. The proper eigenfunctions $ \psi_i$ of
the original Hamiltonian $h_0$ are known exactly
: $ h_0 \psi_i = E_i \psi_i $. Any such operator
$L^{(N)}$ can always be presented in the form
known as Crum-Krein formula \cite{crum} \beq
\displaystyle{L^{(N)} = W^{-1} \left(
u_1,u_2,\cdots,u_N \right) \left|
\begin{array}{cccc} u_1 & u_2 & \cdots & 1 \\
u_1 ^{\prime} & u_2 ^{\prime} & \cdots & \frac{d}{dx} \\
\vdots & \vdots & \ddots & \vdots \\
u_1 ^{(N)} & u_2 ^{(N)} & \cdots & \frac{d^N}{dx^N} \\
\end{array} \right|} \eeq
where $ W \left( u_1,u_2,\cdots,u_N \right) $ stands for the usual
symbol for the Wronskian of the functions $ u_1, u_2, \ldots, u_N
$. The functions $u_i, \ (i=1,2,\ldots,N) $ called the
transformation functions are eigenfunctions of $h_0$, $h_0 u_i =
\alpha _i u_i$, and they need not necessarily satisfy any physical
boundary condition. The final potential has the form \beq V_N (x)
= V(x) - 2\frac{d^2}{dx^2} \ln W(u_1, u_2, \ldots, u_N)
\label{potential} \eeq and will be free of singularities whenever
the Wronskian is nodeless, which in turn, requires that only
consecutive eigenfunctions of $h_0$ have to be considered
\cite{samsonov}. The eigenfunctions $ \psi (x) $ and $ \widetilde{
\psi} (x) $ of $h_0$ and $h_N$ are connected by the intertwiners
$L^{(N)}$ and $L^{(N) \dagger} $ as : \beq \widetilde{\psi} _i (x)
= \displaystyle{L^{(N)} \psi _i (x) =
\f{W_{j,j+1,\cdots,j+N,i}(x)}{W_{j,j+1,\cdots,j+N} (x)} } \eeq
where $ W_{j,j+1,\cdots,j+N,i}(x) $ and $ W_{j,j+1,\cdots,j+N}
(x)$ are the Wronskians of the eigenfunctions of $h_0$ associated
with the corresponding subindices. Thus if $ \psi _i (x) $ is an
eigenfunction of $h_0$ with energy $E_i$, then $ \widetilde{\psi}
_i (x) $ is an eigenfunction of $h_N$ with the same energy $E_i$.
Evidently \beq L^{(N)} \psi _i = 0, \ \ \ \ \ \ \ i=1,2,,\cdots,N
\eeq However, for energies $E_i ( i = 1,2, \cdots N)$, the
corresponding eigenfunctions of $h_N$ $$ \widetilde{\psi} (x) ~~
\a ~~  \f{\psi (x)}{W_{j,j+1,\cdots,j+N} (x)}  $$ have growing
asymptotics at both infinities. Consequently, these are not
physically acceptable solutions of $h_N$, and the corresponding
eigenvalues $E_i ( i = 1,2, \cdots N)$ are excluded from the
spectrum of $h_N$.  Thus \beq h_N \widetilde{\psi}_E = E
\widetilde{\psi}_E \eeq with the exception of the levels $ E =
E_i, i= 1, 2, \cdots, N $, which will be absent in the spectrum of
the new Hamiltonian $h_N$, as the corresponding eigenfunctions are
not square integrable.

It has already been shown \cite{bagrov} that the operator
$L^{(N)}$ can always be presented as a product of $N$ first order
Darboux transformation operators between every two Hamiltonians
$h_0, h_1, \ldots, h_N$ : \beq L^{(N)} = L_N L_{N-1} \cdots L_1
\qquad  h_p L_p = L_p h_{p-1} \qquad p = 1,2,\cdots,N \eeq We note
that the final Hamiltonian  $h_N$ is Hermitian, although some of
the intermediate Hamiltonians $h_i$ could be unphysical, e.g.,
their associated potentials might contain extra singularities that
were not present in the initial one. The supercharges $Q_N$ and
$Q_N ^{\dagger}$ are constructed as \beq Q_N = \left(%
\begin{array}{cc}
  0 & L^{(N)} \\
  0 & 0 \\
\end{array}%
\right), \ \ \ \ \ \ \ \ \ \ \ \ \ \ Q_N ^{\dagger} = \left(%
\begin{array}{cc}
  0 & 0 \\
  L^{(N) \dagger} & 0 \\
\end{array}%
\right) \eeq Evidently, $Q_N$ and $Q_N
^{\dagger}$ are nilpotent \beq \lt\{ Q_N, Q_N
\rt\} = \lt\{ Q_N ^{\dagger}, Q_N ^{\dagger}
\rt\} =0 \eeq The super Hamiltonian
\beq H_N = \left(%
\begin{array}{cc}
  h_0 & 0 \\
  0 & h_N  \\
\end{array}%
\right) \eeq satisfies the relations \beq \lt[ Q_N, H_N \rt] =
\lt[ Q_N ^{\dagger} , H_N \rt] = 0 \eeq The anticommutator can be
generally expressed by a $N$-th order polynomial $\cal{P_N}$ of
the Hamiltonian $H_N$ \beq  {\cal{H}}_N = \displaystyle \lt\{ Q_N
^{\dagger}, Q_N \rt\}  = \left(%
\begin{array}{cc}
  L^{(N) \dagger} L^{(N)} & 0 \\
  0 & L^{(N)} L^{(N) \dagger}  \\
\end{array}%
\right) = \displaystyle \prod _{k=1} ^N \lt( H_N - \a _k {\cal{I}}
\rt) \label{mother} \eeq where $\cal{I}$ is the $2 \times 2$ unit
matrix, and \beq L^{(N) \dagger} L^{(N)} = \displaystyle \prod
_{k=1} ^N \lt( h_0 - \a _k \rt) \eeq \beq L^{(N)} L^{(N) \dagger}
= \displaystyle \prod _{k=1} ^N \lt( h_N - \a _k \rt) \eeq Since
the right hand side of (\ref{mother}) is a polynomial in $H_N$, it
is called nonlinear SUSY or $N$-fold SUSY. The operator
${\cal{H}_N}$ is termed as the {\it Mother Hamiltonian} and
satisfies the commutation relations \cite{samsonov} \beq \lt[ Q_N,
{\cal{H}}_N \rt] = \lt[ Q_N ^{\dagger} , {\cal{H}}_N \rt] = 0 \eeq
For $N =1$, $N$-fold SUSY reduces
to standard SUSY. \\
The most widely studied higher order SUSY is for $N=2$
\cite{fernan,samsonov}, where the formalism reduces to : \beq L^{(2)} = L_2
L_1 \eeq where \beq L_1 = - \partial _x + (\ln u_1 )^{\prime}, \ \
\ \ \ \ L_2 = - \partial _x + (\ln v) ^{\prime}, \ \ \ v = L_1 u_2
\eeq and the isospectral potential turns out to be \beq
\widetilde{V}_2 (x) = V(x) - 2 \f{d^2}{dx^2} \ln W_{j,j+1} (x)
\eeq

\section{Non linear pseudo-SUSY for non Hermitian Hamiltonians}

In this section we extend the concept of nonlinear or $N$-fold
supersymmetry to non Hermitian quantum mechanics. Though the
Darboux algorithm and (nonlinear) supersymmetric quantum mechanics
are equivalent for Hermitian Hamiltonians, the situation is
different for non Hermitian Hamiltonians. However, intertwining
operators $A^{(N)}$ and $B^{(N)}$ can still be constructed with
the help of Darboux transformation. Analogous to the case of
Hermitian quantum mechanics, it will be shown that once a non
Hermitian Schr\"{o}dinger potential $V(x)$ is exactly solvable,
one can construct an isospectral partner $ \widetilde{V}_N (x) $
from (\ref{potential}) \beq \widetilde{V}_N (x) = V(x) -
2\frac{d^2}{dx^2} \ln W(u_1, u_2, \ldots, u_N) \label{complex}\eeq
where $W$ stands for the usual symbol for the Wronskian of the
functions $ u_1, u_2, \ldots, u_N $, which are eigenfunctions of
$h_0$, $h_0 u_i = \alpha _i u_i$. As before the functions $u_i(x)$
may be just formal eigenfunctions. Our aim will be to study the
spectrum of the new Hamiltonian in detail, to investigate the
nature of the potential and the eigenfunctions, and to determine
the symmetry which connects the original Hamiltonian $h_0$ and the
transformed one $h_N$. For this purpose, we look for two
intertwining operators $A^{(N)}$ and $B^{(N)}$ such that \beq
A^{(N)} h_0 = h_N A^{(N)}, \ \ \ \ \ \ \ h_0 B^{(N)} = B^{(N)} h_N
\label{noninter} \eeq where $h_0$ and $h_N$ are no longer
self-adjoint operators ($h_{0,(N)} \neq h_{0,(N)} ^{\dagger}$) ;
on the contrary, to ensure the reality of the spectrum, they are
$\eta$ pseudo Hermitian \beq \eta h_{0,(N)} \eta ^{-1} =
h_{0,(N)}^{\dagger} \label{pseudo} \eeq where $\eta$ is a linear,
invertible, Hermitian operator. However, the choice of $\eta$ is
not unique. For ${\cal{PT}}$ invariant potentials, a simple
representation of $\eta$ may be given by the parity operator :
\beq \displaystyle {\eta = {\cal P},  \qquad {\cal P} f(x) =
f(-x)} \label{parity} \eeq It follows that for real potentials,
(\ref{parity}) leads to $\eta = 1$ so that $B^{(N)} = A^{(N)
\dagger}$, thus reproducing the standard result of supersymmetry.

\vs{.2cm}

It follows from equations (\ref{noninter}) and (\ref{pseudo}) that
the operators $A^{(N)}$ and $B^{(N)}$ are pseudo-adjoint : \beq
B^{(N)} = A^{(N) \#} = \eta ^{-1} A^{(N) \dagger} \eta \eeq
Considering first order Darboux transformation between every two
juxtaposed Hamiltonians $h_0, h_1, \ldots, h_N$, each pair
intertwined by first order operators $L_k ~ (k=1,2,\cdots,N) $
\beq h_k L_k = L_k h_{k-1} \qquad k = 1,2,3, \cdots ,N \eeq \beq
L_k ^{\#} h_k = h_{k-1} L_k ^{\#} \qquad k=1,2,3,\cdots, N \eeq
where \beq L_k ^{\#} = \eta ^{-1} L_k \eta \eeq then, analogous to
the Hermitian case, the final Hamiltonian $h_N$ is found to be
related to the initial (or starting) Hamiltonian $h_0$ through
\beq h_N = L_N  L_{N-1} \cdots L_2  L_1 ~~ h_0 ~~ L_1 ^{\#} L_2
^{\#} \cdots L_N ^{\#} \eeq so that the operator $A^{(N)}$ can be
represented as a product of the $N$ first order Darboux
transformations \beq A^{(N)} = L_N L_{N-1} \ldots L_2 L_1 \eeq
with its pseudo-adjoint \beq B^{(N)} = A^{(N) \#} = \eta ^{-1} L_1
L_2 \ldots L_N \eta = L_1 ^{\#} \ldots L_{N-1} ^{\#} L_N ^{\#}
\eeq It is worth mentioning here that in contrast to Hermitian
quantum mechanics, all the intermediate Hamiltonians $h_k$ are
physically acceptable as their associated potentials contain no
extra singularities which are not present in the initial potential
$V(x)$. This is essentially because the associated eigenfunctions
do not have nodes on the real line, and they are normalizable in
the sense of equation (\ref{cpt}).

Thus the initial and the transformed Hamiltonians
$h_0$ and $h_N$ are related by {\it non linear
pseudo supersymmetry}.  The super
Hamiltonian of this system consists of the
pseudo supersymmetric pair of Hamiltonians $h_0$
and $h_N$ as
\beq H_N = \left(%
\begin{array}{cc}
  h_0 & 0 \\
  0 & h_N  \\
\end{array}%
\right) \label{hh} \eeq The supercharges
generating this form of pseudo-supersymmetry are
constructed in the following way :
\beq Q_N= \left(%
\begin{array}{cc}
  0 & A^{(N)} \\
  0 & 0 \\
\end{array}%
\right), \ \ \ \ \ \ \ \ \ \ \ \ \ \ Q_N ^{\#} = \eta ^{-1}
Q_N ^{\dagger} \eta = \left(%
\begin{array}{cc}
  0 & 0 \\
  B^{(N)}  & 0 \\
\end{array}%
\right) =
\left(%
\begin{array}{cc}
  0 & 0 \\
  A^{(N) \#}  & 0 \\
\end{array}%
\right) \eeq so that the supercharge $Q_N$ and its adjoint $Q_N
^{\dagger}$ of standard Hermitian quantum mechanics are replaced
by $Q_N$ and its pseudo adjoint $Q_N ^{\#}$ for non Hermitian
Hamiltonians. Obviously, $Q_N$ and $Q_N ^{\#}$ are nilpotent \beq
\lt\{ Q_N, Q_N \rt\} = \lt\{ Q_N ^{\#}, Q_N ^{\#} \rt\} =0 \eeq
and satisfy the following closed algebra : \beq \lt[ Q_N, H_N \rt]
= \lt[ Q_N ^{\#} , H_N \rt] = 0 \eeq \beq {\cal{H}}_N \equiv \lt\{
Q_N ^{\#}, Q_N \rt\}
= \left(%
\begin{array}{cc}
  A^{(N) \#} A^{(N)} & 0 \\
  0 & A^{(N)} A^{(N) \#}  \\
\end{array}%
\right) = \displaystyle \prod _{k=1} ^N \lt( H_N - \a _k {\cal{I}}
\rt)  \eeq i.e., \beq A^{(N) \#} A^{(N)} = \displaystyle \prod
_{k=1} ^N  \lt( h_0 - \alpha _k \rt)  \eeq \beq A^{(N)} A^{(N) \#}
= \displaystyle \prod _{k=1} ^N \lt( h_N - \alpha _k \rt) \eeq and
${\cal{I}}$ is the $2 \times 2$ unit matrix.
 Evidently, if $\psi _i (x)$ is an
eigenfunction of $h_0$ with energy eigenvalue $E_i$, then $
\widetilde{\psi} _i (x) = A^{(N)} \psi _i (x) $  is an
eigenfunction of $h_N$ with the same energy $E_i$. However, for $
i = 1,2,\cdots,N $, \beq \widetilde{\psi} (x) ~~ \alpha ~~ \f{\psi
(x)}{ W \lt( \psi _1, \psi _2, \cdots, \psi _N \rt) } \eeq
Clearly, the eigenfunctions $ \widetilde{\psi} _i (x) ~ ( i = 1,2,
\cdots N) $ of $h_N$ corresponding to the eigenvalues $E_i ~
(i=1,2,\cdots,N)$ grow asymptotically, and so cannot be included
in the set of solutions of $h_N$. Consequently, $E_i ~
(i=1,2,\cdots,N)$  are excluded from the spectrum of $h_N$.

\vs{.3cm}

\noindent Next we note two interesting results which are in
contrast to the Hermitian case:

\vs{.25cm}

\noindent {\bf 1.} ~~ For $\widetilde{V} _N (x)$ to be free of
singularities, the Wronskian $ W \lt( \psi _1, \psi _2, \cdots,
\psi _N \rt)  ~=~ W_{\psi _1, \psi _2, \cdots, \psi _N } (x) $
must be nodeless. In case of Hermitian potentials, this is
guaranteed only when $\psi _i, \ i = 1, 2, \cdots N $ represent
$N$ consecutive eigenfunctions. However, in case of generic non
Hermitian potentials, the eigenfunctions $\psi_n (x), ~ (n=0,1,2,
\cdots) $ have no nodes on the real line. Consequently, the
Wronskian  is free of real singularities for any value of $i,j,k,
\cdots $, and thus can be used to generate a wider class of isospectral
Hamiltonians.

\vs{.25cm}

\noindent {\bf 2.} ~~ The intermediate Hamiltonians are also
physically acceptable, as the corresponding potentials are free of
singularities, for the same reason as given above. For example the
first intertwining gives \beq V_1 (x) = V(x) - 2 \f{d^2}{dx^2} \ln
\psi _i (x) \eeq which is well defined. However, this may not
always be true for Hermitian potentials due to the presence of
additional singularities in $V_1 (x)$, which are not present in
$V(x)$.

\vs{.5cm}

For the sake of simplicity, in the present work we shall restrict
ourselves to second order nonlinear pseudo-supersymmetry. Thus if
an intertwining operator $A = L_2 L_1 $ is constructed from the
two first order Darboux transformation operators $L_1$ and $L_2$,
given by \beq L_1 = -
\partial _x + (\ln u_i )^{\prime}, \ \ \ \ \ \ L_2 = - \partial _x
+ (\ln v) ^{\prime}, \ \ \ v = L_1 u_j \eeq where $u_i$ and $u_j$
are any two eigenfunctions of the non Hermitian Hamiltonian $h_0$,
then the transformed isospectral Hamiltonian \beq h_2  =
-\f{d^2}{dx^2} + \widetilde{V}_{i,j} (x) \eeq  has eigenfunctions
\beq \ba {lcl} \widetilde{\psi} _n (x) &=& \displaystyle \f{ W
\lt( \psi _i, \psi _j, \psi _n \rt) }{ W \lt( \psi _i, \psi _j
\rt) } \\ &=& \displaystyle - E_n \psi _n + E_i \psi _i \f{ W \lt(
\psi _n , \psi _j \rt)}{ W \lt( \psi _i, \psi _j \rt) } + E_j \psi
_j \f{ W \lt( \psi _i , \psi _n \rt)}{ W \lt( \psi _i, \psi _j
\rt) } \ea \label{eigenfn} \eeq where \beq \widetilde{V}_{i,j} (x)
= \displaystyle V(x) - 2 \f{d^2}{dx^2} \ln W \lt( u_i,u_j \rt)
\eeq The mother Hamiltonian $ {\cal{H}} _2 $ is constructed from
the anticommutator by \beq {\cal{H}}_2 = \lt\{ Q_2 ^{\#}, Q_2
\rt\}  = \left(%
\begin{array}{cc}
  A^{\#} A & 0 \\
  0 & A A^{\#}  \\
\end{array}%
\right)  = \lt(%
\ba {cc}
\lt( h_0 - \alpha _1 \cal{I} \rt) \lt( h_0 - \alpha _2
\cal{I} \rt) & 0 \\
0 & \lt( h_2 - \alpha _1 \cal{I} \rt) \lt( h_2 - \alpha _2 \cal{I}
\rt) \\
\ea %
\rt) \eeq where $\cal{I}$ is $2 \times 2$ unit matrix and $H_2$ is
given by (\ref{hh}).

\vs{.3cm}

In the following sections we shall investigate this formalism further with the help of explicit examples.

\section{${\cal{PT}}$ Symmetric Oscillator}

In this section we shall apply our formalism to the well known
example of the ${\cal{PT}}$ symmetric oscillator \cite{znojil}
\beq V(x) = \lt( x - i \eps \rt) ^2 + \f{\alpha ^2 -
\f{1}{4}}{\lt( x - i \eps \rt) ^2} \label{pto} \eeq with
eigenfunctions \beq \psi _{n} (x) = \displaystyle {e^ {- \f{1}{2}
\lt( x - i \eps \rt) ^2} \lt( x - i \eps \rt) ^{-q \alpha +
\f{1}{2} } L_n ^{-q \alpha} \lt( \lt( x - i \eps \rt)^2 \rt) }
\eeq and eigenvalues \beq E_{n} = 4n -2q \alpha + 2 , \qquad n =
0,1,2, \cdots \eeq where $q= \pm 1$ is called the quasi-parity.

In this study we shall restrict ourselves to $N=2$ only. If one
performs Darboux transformations with two eigenfunctions $\psi _i
(x) $ and $ \psi _j (x) $ of the potential $V(x)$, corresponding
to energies $E_i$ and $E_j$ ($i$ and $j$ need not be consecutive),
then the intertwining operators take the form \beq L_1 = -
\f{d}{dx} + \f{\psi _i ^{\pr}}{\psi _i} \ , \qquad \qquad L_1
^{\#} = \f{d}{dx} + \f{\psi _i ^{\pr}}{\psi _i} \eeq \beq L_2 = -
\f{d}{dx} + \f{W_{i,j} ^{\pr}}{W_{i,j}} - \f{\psi _i ^{\pr}}{\psi
_i} \ , \qquad  \qquad L_2 ^{\#} = \f{d}{dx} + \f{W_{i,j}
^{\pr}}{W_{i,j}} - \f{\psi _i ^{\pr}}{\psi _i} \eeq where
$W_{i,j}$ is the usual Wronskian given by \beq W_{i,j} = W(\psi
_i, \psi _j ) = \psi _i (x) \psi _j ^{\pr} (x) - \psi _i ^{\pr}
(x) \psi _j (x) \eeq and $\eta$ has been taken as in
(\ref{parity}). Replacing the intertwining operators $A^{(2)}$
(and $B^{(2)}$) by $A$ (and $B$) for simplicity, we obtain \beq A
~=~ L_2 L_1 ~=~ \f{d^2}{dx^2} + \beta _{i,j} \f{d}{dx} - \beta
_{i,j} \f{\psi _i ^{\pr} }{ \psi _i} - \f{\psi _i ^{\pr \pr} }{
\psi _i} \eeq \beq B ~=~ A^{\#} ~=~ \eta ^{-1} A ^{\dagger} \eta
~=~ \f{d^2}{dx^2} - \beta _{i,j} \f{d}{dx} - \beta _{i,j} \f{\psi
_i ^{\pr} }{ \psi _i} - \f{\psi _i ^{\pr \pr} }{ \psi _i} - \beta
_{i,j} ^{\pr} \eeq where \beq \beta _{i,j} = - \f{W_{i,j} ^{\pr}
}{W_{i,j}} \eeq and $\eta $ is simply the parity operator
${\cal{P}}$ for ${\cal{PT}}$ symmetric potentials. The new exactly
solvable non Hermitian potential, which is isospectral to the
${\cal{PT}}$ symmetric oscillator in (\ref{pto}), is obtained from
\beq \widetilde{V} _{i,j} (x) = V(x) - 2 \f{d^2}{dx^2} \ln W_{i,j}
\label{newpot} \eeq with solutions \beq \widetilde{ \psi } _k (x)
= A \psi _k (x) \eeq Thus for each set $(i,j)$, one obtains two
sets of $\widetilde{V}_{i,j} (x)$ because of the presence of quasi
parity $q$. Obviously, $ \widetilde{\psi} _k (x) = 0$ for $k=i,j$.
Thus the new potential so constructed, in (\ref{newpot}) above,
has all the eigenenergies of the original ${\cal{PT}}$ symmetric
oscillator except for the levels $i,j$, which are missing from the
spectrum of (\ref{newpot}).

\vs{.2cm}

\noindent For the simplicity of calculations we shall now construct
and examine some potentials using low values of $i$ and $j$ in
further detail.

\vs{.5cm}

\subsection{New potential for $i=1$, $j=2$}

\vs{.2cm}

\noindent Applying the above formalism with the two eigenstates
$\psi _1 (x)$ and $ \psi _2 (x)$, of the potential in (\ref{pto}),
the Wronskian is found to be \beq \displaystyle W \lt( \psi _1,
\psi _2 \rt)  = \displaystyle c_{12} e ^{ - \lt( x - i \eps \rt)
^2} \lt( x - i \eps \rt)^{2 - 2 q \a} g \eeq where $c_{12}$ is
some real constant and \beq g = \displaystyle \lt(1 - q \a \rt)
\lt( 2 - q \a \rt) - 2 \lt( 1 - q \a \rt) \lt( x - i \eps \rt)^2 +
\lt( x - i \eps \rt)^4  \eeq The intertwining operators $A$ and
$A^{\#}$ are obtained from $ A = L_2 L_1 $, $ A^{\#} = L_1 ^{\#}
L_2 ^{\#} $, where \beq \ba {lcl} L_1
&=& \displaystyle - \f{d}{dx} + \f{\psi _1 ^{\pr}}{ \psi _1} \\
&=& \displaystyle - \f{d}{dx} - \lt( x - i \eps \rt) + \f{ -q \a +
\f{1}{2} }{ \lt( x - i \eps \rt)} - \f{2 \lt( x - i \eps \rt) }{ 1
- q \a - \lt( x - i \eps \rt)^2 } \ea \eeq \beq \ba {lcl} L_2 &=&
\displaystyle - \f{d}{dx} + \f{d}{dx} \ln (L_1 \psi _2 )\\
&=& \displaystyle - \f{d}{dx} - \lt( x - i \eps \rt) + \f{ - q \a
+ \f{3}{2} }{ \lt( x - i \eps \rt)} + \f{2 \lt( x - i \eps \rt) }{
1 - q \a - \lt( x - i \eps \rt) ^2 } + \f{g^{\pr}}{g} \ea \eeq so
that \beq \ba {lcl} A &=& \displaystyle \f{d^2}{dx ^2} - \lt\{ -2
\lt( x - i \eps \rt) + \f{2 \lt( 1 - q \a \rt) }{ \lt( x - i \eps
\rt) } + \f{ g ^{ \pr }}{g} \rt\} \f{d}{dx} + \lt( x - i \eps
\rt)^2 + \f{ \lt( -q \a + \f{1}{2} \rt) \lt( - q \a + \f{5}{2}
\rt) }{ \lt( x - i \eps \rt)^2 } \\ \\ &~& \displaystyle + ~ 2 q
\a - 1 + \lt\{ - \lt( x - i \eps \rt) + \f{- q \a + \f{1}{2}}{
\lt( x - i \eps \rt) } - \f{ 2 \lt( x - i \eps \rt)}{ 1 - q \a -
\lt( x - i \eps \rt)^2 } \rt\} \f{g ^{\pr}}{g} \ea \eeq \beq \ba
{lcl}  A^{\#}   &=& \displaystyle \f{d^2}{dx ^2} + \lt\{ -2 \lt( x
- i \eps \rt) + \f{2 \lt( 1 - q \a \rt) }{ \lt( x - i \eps \rt) }
+ \f{ g ^{ \pr }}{g} \rt\} \f{d}{dx} + \lt( x - i \eps \rt)^2 +
\f{ \lt( -q \a + \f{3}{2}
\rt) \lt( - q \a - \f{1}{2} \rt) }{ \lt( x - i \eps \rt)^2 } \\ \\
&~& \displaystyle + ~ 2 q \a - 3 + \lt\{ - \lt( x - i \eps \rt) +
\f{- q \a + \f{1}{2}}{ \lt( x - i \eps \rt) } - \f{ 2 \lt( x - i
\eps \rt)}{ 1 - q \a - \lt( x - i \eps \rt)^2 } \rt\} \f{g
^{\pr}}{g} + \f{ g ^{\pr \pr}}{g} - \lt( \f{g ^{\pr}}{g} \rt) ^2
\ea \eeq

Applying equation (\ref{newpot}), the new potential isospectral to
the one in (\ref{pto}) except for the states corresponding to $
\psi _1 (x)$ and $ \psi _2 (x) $, comes out as \beq \widetilde{V}
_{1,2} (x) = \displaystyle \lt( x - i \eps \rt) ^2 + \f{ \lt( -q
\a + \f{3}{2} \rt) \lt( -q \a + \f{5}{2} \rt) }{ \lt( x - i \eps
\rt)^2} - 2 \f{ g ^{ \pr \pr}}{g} + 2 \lt( \f{g ^{ \pr} }{g}
\rt)^2 + 4 \label{v12}\eeq which has solutions \beq
\widetilde{\psi} _n (x) = \displaystyle - E_{n+2} \psi _{n+2} +
E_1 \psi _1 \f{ W \lt( \psi _{n+2} , \psi _2 \rt)}{ W \lt( \psi
_1, \psi _2 \rt) } + E_2 \psi _2 \f{ W \lt( \psi _1 , \psi _{n+2}
\rt)}{ W \lt( \psi _1, \psi _2 \rt) } \eeq with energy eigenvalues
\beq \widetilde{E} _{n} = E_{n+2} = 4n + 10 - 2 q \a \ , \qquad
\qquad  n = 1, 2, 3, \cdots \eeq The ground state is given by \beq
\widetilde{ \psi} _0 (x) = \displaystyle e^{- \f{1}{2} (x - i \eps
)^2} (x - i \eps )^{ -q \a + \f{1}{2}} \lt\{ B_1 + \f{B_2 ( x - i
\eps )^2 }{g} \rt\} \eeq with eigenvalue \beq \widetilde{E} _{0} =
E_{0} = 2 - 2 q \a \eeq where $B_1$ and $B_2$ are some $x,\eps$
independent constants. Thus the energies $E_1$
and $E_2$ of $V(x)$ are absent in the spectrum of
$\widetilde{V}_{1,2} (x)$. It can be verified that the
eigenfunctions $\widetilde{ \psi } $ are also ${\cal{PT}}$
invariant, and can be normalized using (\ref{cpt}). Furthermore,
the supercharges $Q_2$ and $Q_2 ^{\#} $, generated from the
operators $A$ and $A^{\#}$, satisfy the following algebra : \beq
{\cal{H}}_2 = \displaystyle \lt\{ Q_2, Q^{\#} _2 \rt\} =  H_2 ^2 -
4(4 -  q \a ) H_2 + ( 6 - 2 q \a ) (10 - 2 q \a ) \eeq where $ H_2
$ is given by (\ref{hh}). The intermediate potential given by \beq
\ba {lcl}
V_1 (x) &=& \displaystyle V(x) - 2 \f{d^2}{dx^2} \ln \psi_1 (x) \\
\\ &=& \displaystyle \lt( x - i \eps \rt)^2 + \f{\lt( - q \a +
\f{1}{2} \rt) \lt( - q \a + \f{3}{2} \rt) }{ \lt( x - i \eps
\rt)^2 } + \f{12}{ 1 - q \a - \lt( x - i \eps \rt)^2 } - \f{ 8
\lt( 1 - q \a \rt) }{ \lt\{ 1 - q \a - \lt( x - i \eps \rt)^2
\rt\} ^2 } + 4 \ea \label{v1}\eeq does not have any singularity on
the real line, and hence is physically acceptable as well. By
arguments similar to those given above, its ground state
eigenfunction is given by : \beq \phi _ 0 = \displaystyle \f{1}{ 1
- q \a - (x - i \eps ) ^2 } ~ e^{ - \f{1}{2} ( x - i \eps ) ^2} (
x - i \eps ) ^{ -q \a + \f{3}{2} } \eeq with energy \beq e_{0} =
E_{0} = 2 - 2 q \a \eeq and the excited states \beq \phi _n =
\displaystyle \f{ W \lt( \psi _{n+1}, \psi _1 \rt) }{ \psi _1}
\eeq with corresponding energies \beq e _{n} = E_{n+1} = 4n + 6 -
2 q \a , \qquad n = 1,2,3, \cdots \eeq It is easy to observe that
applying (\ref{pt}), both the intermediate and the final
potentials (as well as their eigenfunctions) satisfy
(\ref{ptsym}), and hence are ${\cal{PT}}$ invariant, having real
spectra.

\vs{1cm}

\subsection{New Potentials for $i=0$, $j=2$}

\vs{.2cm}

\noindent In a similar manner, the expressions for the different
quantities are obtained as follows : \beq \displaystyle W \lt(
\psi _ 0, \psi _2 \rt)  = \displaystyle c_{02} e^{- \lt( x - i
\eps \rt) ^2} \lt( x - i \eps \rt) ^{-2q \a + 2} \lt\{ \f{ \lt( x
- i \eps \rt)^2 }{2-q \a} -1 \rt\} \eeq with $c_{02}$ some real
constant \beq L_1 = \displaystyle - \f{d}{dx} - \lt( x - i \eps
\rt) + \f{ \lt( -q \a + \f{1}{2} \rt) }{ \lt( x - i \eps \rt) }
\eeq \beq L_2 = \displaystyle - \f{d}{dx} - \lt( x - i \eps \rt) +
\f{ \lt( -q \a + \f{3}{2} \rt)}{\lt( x - i \eps \rt)} + \f{ 2 \lt(
x - i \eps \rt)}{\lt[ \lt( x - i \eps \rt)^2 - \lt(2 - q \a \rt)
\rt] } \eeq \beq \ba{lcl} A &=& \displaystyle \f{d^2}{dx^2} + 2
\lt\{ \lt( x - i \eps \rt) + \f{ \lt( q \a -1 \rt) }{ \lt( x - i
\eps \rt) } - \f{ \lt( x - i \eps \rt)}{ \lt( x - i \eps \rt)^2 -
\lt( 2 - q \a \rt) } \rt\} \f{d}{dx} + \lt( x - i \eps \rt)^2 \\
\\ &~& \displaystyle + \f{ \lt( -q \a + \f{1}{2} \rt) \lt( - q \a
+ \f{5}{2} \rt) }{ \lt( x - i \eps \rt)^2 } - \f{3}{ \lt( x - i
\eps \rt)^2 - \lt( 2 - q \a \rt) } + 2 q \a - 3  \ea \eeq \beq
\ba{lcl} A^{\#} &=& \displaystyle \f{d^2}{dx^2} - 2 \lt\{ \lt( x -
i \eps \rt) + \f{ \lt( q \a -1 \rt) }{ \lt( x - i \eps \rt) } -
\f{ \lt( x - i \eps \rt)}{ \lt( x - i \eps \rt)^2 - \lt( 2 - q \a
\rt) } \rt\} \f{d}{dx} \\ \\ &~& \displaystyle + \lt( x - i \eps
\rt)^2 + \f{ \lt( -q \a - \f{1}{2} \rt) \lt( - q \a + \f{3}{2}
\rt) }{ \lt( x - i \eps \rt)^2 } - \f{5}{ \lt\{ \lt( x
- i \eps \rt)^2 - \lt( 2 - q \a \rt) \rt\} } \\
\\ &~& \displaystyle  - \f{4 (-q \a + 2 )}{ \lt\{
\lt( x - i \eps \rt)^2 - \lt( 2 - q \a \rt) \rt\}
^2 } + 2 q \a - 5 \ea \eeq
The new potential \beq
\ba {lcl} \widetilde{V} _{0,2} (x) &=&
\displaystyle \lt( x - i \eps \rt)^2 + \f{\s \lt(
\s - 1 \rt) }{ \lt( x - i \eps \rt)^2 } + \f{4}{
\lt( x - i \eps \rt)^2 - \lt( 2 - q \a \rt) } \\
\\ &~& + \displaystyle \f{ 8 \lt( 2 - q \a \rt)
}{ \lt\{ \lt( x - i \eps \rt)^2 - \lt( 2 - q \a \rt) \rt\} ^2 } +
4 \ea \label{new} \eeq where \beq \s = - q \a + \f{5}{2} \eeq is
totally different from the initial potential of the ${\cal{PT}}$
symmetric oscillator, yet shares the same spectrum except for the
states $n=0,2$ of the original potential, which are missing in the
partner. \\
The ground state wave function of the Hamiltonian in (\ref{new})
is given by \beq \widetilde{\psi} _0 (x) = \displaystyle \lt\{ A_1
\lt( x - i \eps \rt) ^2 + A_2 + \f{A_3}{ \lt( x - i \eps \rt) ^2 -
\lt( 2 - q \a \rt) } \rt\} e^{ - \f{1}{2} \lt( x - i \eps \rt) ^2}
\lt( x - i \eps \rt) ^{ - q \a + \f{1}{2} } \eeq with ground state
energy \beq \widetilde{E}_0 = E_1 = 6 - 2 q \a \eeq  where
$A_1,A_2,A_3$ are $x$-independent constants, while the excited
states are obtained from (\ref{eigenfn}) \beq
\ba {lcl} \widetilde{ \psi} _n &=& \displaystyle A \psi _{n+2} \\
&=& \displaystyle -E_{n+2} \psi _{n+2} + E_0 \psi _0 \f{ W \lt(
\psi _{n+2}, \psi _2 \rt) }{ W \lt( \psi _0, \psi _2 \rt) } + E_2
\psi _2 \f{ W \lt( \psi _0, \psi _{n+2} \rt) }{ W \lt( \psi _0,
\psi _2 \rt) } \ea \eeq with energies \beq \widetilde{E}_n =
E_{n+2} = 4n + 10 - 2 q \a \ , \qquad \qquad n=1,2,\cdots \eeq It
can also be verified that eigenfunctions $ \widetilde{\psi} _n (x)
$ have correct asymptotic behaviour and are also ${\cal{PT}}$
invariant. Consequently, they also satisfy equation (\ref{cpt}).
The intermediate potential is given by \beq V_1 (x) =
\displaystyle \lt( x - i \eps \rt)^2 + \f{\lt( - q \a + \f{1}{2}
\rt) \lt( -q \a + \f{3}{2} \rt) }{\lt( x - i \eps \rt)^2 } + 2
\label{inter} \eeq which is also physically acceptable. By
arguments similar to those given above, its ground state
eigenfunction is given by \beq \phi _0 = \displaystyle e ^{ -
\f{1}{2} (x - i \eps )^2} \lt( x - i \eps \rt) ^{-q \a + \f{3}{2}
} \eeq with energy \beq e_0 = 6 - 2 q \a \eeq and  excited states
\beq \phi _n = \displaystyle \f{ W \lt( \psi _{n+1} , \psi _0 \rt)
}{ \psi _0 } \eeq with energies \beq e _{n} = E_{n+1} = 4n + 6 - 2
q \a \ , \qquad \qquad n = 1,2,3, \cdots \eeq  Once again, both
the intermediate and the final potentials (as well as their
eigenfunctions) are ${\cal{PT}}$ invariant, having real
spectra.\\
The supercharges $Q_2$ and $Q_2 ^{\dagger} $ generated from the
intertwining operators $A$ and $A^{\#}$ can be shown to satisfy
the following algebra : \beq {\cal{H}}_2 = \displaystyle \lt\{
Q_2, Q^{\#} _2 \rt\} =  H_2 ^2 - 4(3 -  q \a ) H_2 + ( 2 - q \a )
(10 - 2 q \a ) \eeq where $ H_2 $ is given by (\ref{hh}).

We note that the potentials obtained in this section are unique in
the sense that they do not have any counterpart in standard
quantum mechanics (i.e., in the Hermitian case). .

\vs{1cm}

\section{${\cal{PT}}$ symmetric Scarf II potential}

We note that the generalised oscillator problem considered in the
last section was made non Hermitian by an imaginary displacement
of the coordinate variable $x$. However, there are other methods
of constructing non Hermitian models. To see how the formalism
described in section 3 works with such models, in this section we
shall study an example, viz., the ${\cal{PT}}$ symmetric non
Hermitian Scarf II potential, which has been ${\cal{PT}}$
symmetrized in a different way. This exactly solvable potential,
given by \beq V(x) = - \lambda ~ sech ^2 x - i \mu ~ sech ~x ~
tanh ~x, ~~~~~~~~~~~~~~~ \lambda > 0, ~ \mu \neq 0 \label{scarf2}
\eeq has a discrete spectrum that admits both real as well as
complex conjugate energies, depending on the relative strengths of
its parameters $\lambda$ and $\mu$. For $ | \mu | \leq \lambda +
\f{1}{4} $, the system possesses a real and discrete bound state
spectrum, whereas for $ | \mu | > \lambda + \f{1}{4} $, the system
exhibits spontaneous ${\cal{PT}}$ symmetry breaking, with complex
conjugate pairs of energies. The normalized wave functions for
this potential are well known, being given by \cite{jpa,zafar}
\beq \psi _n (x) = \f{ \Gamma \lt( n - 2p + \f{1}{2} \rt) }{ n!
\Gamma \lt( \f{1}{2} - 2p \rt)}~ z^{-p}~(z^*)^{-q} ~ P_n ^{-2p -
\f{1}{2}, ~ -2q - \f{1}{2}} ( i~sinh ~x) \label{wavescarf} \eeq
where $P_n ^{\alpha, \beta}$ are the Jacobi polynomials
\cite{handbook}: \beq P_n ^{\alpha,\beta} (i~sinh ~x) = \f{\Gamma
(n+ \a +1)}{\Gamma (n+1) \Gamma (\a +1)} ~ F \lt( -n, n + \alpha +
\beta + 1; \alpha + 1; z \rt) \eeq and \beq z = \f{1-i~sinh~x}{2}
\eeq \beq p = - \f{1}{4} \pm \f{1}{2} \sqrt{\f{1}{4} + \lambda +
\mu } = - \f{1}{4} \pm \label{p} \f{t}{2} \eeq \beq q = - \f{1}{4}
\pm \f{1}{2} \sqrt{\f{1}{4} + \lambda - \mu } =  - \f{1}{4} \pm
\f{s}{2}\label{q} \eeq  However, for normalization of the wave
functions, only the positive  sign is allowed in $p$. The energy
spectrum \beq E_n = - \lt( n -p - q \rt)^2, ~ ~ ~ ~ ~ n=0, 1,
2,... < \lt( \f{s+t-1}{2} \rt) \eeq is real and bound for $ |\mu|
\leq \lambda + \f{1}{4} $ , i.e., for real $p$ and $q$, with two
towers characterized by the two values of $q$.

\vs{.5cm}

If the formalism developed above is applied to this example for
$N=2$, with states $\psi _0 (x) $ and $ \psi _2 (x) $, then the
Wronskian is calculated to be \beq W \lt( \psi_0, \psi_2 \rt) =
\displaystyle \lt( 1 - i ~ \sinh x \rt) ^{-2p} \lt( 1 + i ~ \sinh
x \rt) ^{-2q} \cosh x ~ \lt\{ -i (p-q)  + \lt( p + q - \f{3}{2}
\rt) \sinh x \rt\} \eeq and the intertwining operators $A$ and
$A^{\#}$ are given by\beq A = L_2 L_1 \qquad A^{\#} = L_1 ^{\#}
L_2 ^{\#} \eeq where $L_1$ and $L_2$ take the form \beq \ba {lcl}
L_1 &=& \displaystyle
-\f{d}{dx} + \f{\psi_0 ^{\pr}}{\psi_0} \\ \\
&=& \displaystyle \f{d}{dx} + i(p-q) sech ~x - (p+q) tanh ~x \ea
\eeq \beq \ba {lcl} L_2 &=& \displaystyle  -\f{d}{dx} + \f{
W_{0,2} ^{\pr} }{ W_{0,2} }  - \f{\psi_0 ^{\pr}}{\psi_0} \\ \\ &=&
\displaystyle \f{d}{dx} + i(p-q) sech ~x - (p+q) \tanh ~x \\ \\
&~& \displaystyle +\f{ \lt(-p-q+\f{3}{2} \rt) +i \lt( p - q \rt)
\sinh ~x + \lt( -2p -2q + 3 \rt) \sinh ^2 x  }{ i \lt( p-q \rt)
\cosh ~ x + \lt( -p-q+ \f{3}{2} \rt) \sinh ~x ~~ \cosh ~x } \ea
\eeq Now using (\ref{complex}) the new potential is found to be
\beq \widetilde{V} _{0,2} (x) = \displaystyle -
\widetilde{\lambda}~ sech ^2 x - i ~ \widetilde{\mu} ~ sech ~x ~~
tanh ~x - 2 \lt( \f{ \s ^2 ~ sech ^2 x - i ~ \rho ~ \s ~ sech ~ x ~~
tanh ~ x }{ \lt( \rho ~ sech ~x - i ~ \s ~ tanh ~ x \rt) ^2 } \rt)
\label{scarfnew} \eeq where \beq \widetilde{\lambda} =
\displaystyle \lambda - 4p - 4q +2 \eeq \beq \widetilde{\mu} =
\displaystyle \mu - 4p + 4q \eeq \beq \lambda = \displaystyle 2
\lt( p^2 + q^2 \rt) + \lt( p + q \rt) \eeq \beq \mu =
\displaystyle 2 \lt( p ^2 - q^2 \rt) + \lt( p-q \rt) \eeq \beq
\rho = p-q \eeq \beq \s = \displaystyle \lt( -p-q + \f{3}{2} \rt)
\eeq Once again, the final potential $ \widetilde{V} _{0,2} (x)$
is also ${\cal{PT}}$ invariant. The eigen functions are obtained
from (\ref{eigenfn}), with the ground state as \beq \widetilde{
\psi } _0 = \displaystyle \lt( E_0 - E_1 \rt) \psi _1 + \lt( E_2 -
E_0 \rt) \psi _2 \f{ P_1 ^{\pr} }{ P_2 ^{\pr}} \eeq and excited
states \beq \widetilde{\psi} _n = \displaystyle \lt( E_0 - E_{n+2}
\rt) \psi _{n+2} + \lt( E_2 - E_0 \rt) \psi _2 \f{P_{n+2} ^{\pr}
}{ P_2 ^{\pr}} \eeq where $P_n$ denotes the Jacobi polynomial $
P_n ^{-2p - \f{1}{2}, ~ -2q - \f{1}{2}} ( i~sinh ~x) $ and $P_n
^{\pr}$ denotes its derivative with respect to $x$. It can be
shown that for $ |\mu| \leq \lambda + \f{1}{4} $, the
wavefunctions $\widetilde{ \psi} $ are also ${\cal{PT}}$
invariant, and can be normalized following (\ref{cpt}). The new
potential $ \widetilde{V}_{0,2} (x) $ has real bound state
spectrum given by \beq \widetilde{E}_0 = \displaystyle - \lt( 1 -
p - q \rt) ^2 \eeq \beq \widetilde{E}_n = \displaystyle - \lt( n+2
- p - q \rt) ^2,  \qquad n= 1 , 2, \cdots, < \lt( \f{s+t-5}{2}
\rt) \eeq and the algebra satisfied by the supercharges turns out
to be \beq {\cal{H}}_2 = \displaystyle \lt\{ Q_2, Q^{\#} _2 \rt\}
= \displaystyle  H_2 ^2 + \lt( 2 -2p -2q \rt) H_2 + \lt( p + q
\rt)
\lt( p + q - 2 \rt) \eeq where $ H_2 $ is given by (\ref{hh}). \\
The intermediate potential takes the form \beq V_1 (x) = -
\widetilde{v}_1 ~ sech ^2 x - i ~ \widetilde{v}_2 ~ sech ~x ~~
tanh ~x \label{scarfinter}\eeq where \beq \widetilde{v}_1 =
\displaystyle \lambda - 2 \lt( p + q \rt) \eeq  \beq
\widetilde{v}_2 = \displaystyle \mu - 2 \lt( p - q \rt) \eeq  with
eigenfunctions \beq \phi _n = \displaystyle \f{ W \lt( \psi _{n+1}
, \psi _0 \rt) }{ \psi _0 } \label{phi}\eeq and the corresponding energies
\beq e_n = E_{n+1} = -(n+1-p-q)^2~, \qquad \qquad n = 0 , 1,
\cdots, < \lt( \f{s+t-3}{2} \rt)\eeq Thus $ V_1 (x)$ and the corresponding wave functions (\ref{phi}) are also
physically acceptable as well as ${\cal{PT}}$ invariant.

\section{Conclusions}
In this article we have suggested an application of higher order
Darboux algorithm to non Hermitian $\cal{PT}$ symmetric
potentials. For the sake of definiteness the method has been
applied to two specific potentials, namely, the generalised
oscillator and the Scarf II potentials and a number of new
potentials having nearly the same spectrum as the original ones
have been obtained. It may be noted that in each of these cases,
starting from a $\cal{PT}$ symmetric potential we have obtained
new potentials which are again $\cal{PT}$ symmetric. In other
words the higher order Darboux algorithm does not induce
spontaneous $\cal{PT}$ symmetry breaking. Among the different
cases considered here the one involving non consecutive levels
deserves special mention. The potentials thus obtained have no
Hermitian analogues. Also the intermediate potentials in all the
cases are perfectly well behaved since the Darboux algorithm does
not introduce any new singularity or break $\cal{PT}$ symmetry.
Furthermore it has been shown that the symmetry underlying the
original and the new potentials is a fusion of {\it nonlinear
SUSY} and $\cal{PT}$ symmetry which we call {\it nonlinear pseudo
supersymmetry}. Finally we note that analogous to the study of breaking {\it N} fold supersymmetry \cite{tan}, it would be of interest to examine breaking of this new symmetry.

\section*{Acknowledgment}
One of the authors (A.S.) thanks the Council of Scientific \&
Industrial Research, India, for financial assistance.
\pb

\pb

\ed

For a visual representation we have plotted the real and imaginary
parts of the potentials in (\ref{pto}), (\ref{v12}), (\ref{v1}),
(\ref{new}) and (\ref{inter}) in figures (1) - (4) for $q=1$ (for
$q=-1$ the changes are only quantitative and the qualitative
features remain the same). From the figures we find that all the
potentials exhibit the expected behaviour

As before we have plotted the real and imaginary parts of the
potentials (\ref{scarf2}), (\ref{scarfnew}) and (\ref{scarfinter})
in figures 5 and 6. From the figures we find that all the
potentials have acceptable behaviour.

\section*{Figure Captions}

\vs{1cm}

\noindent {\bf Fig. 1} : Plot of real parts of the potentials in
equations (65) (dotted curve), (82) (solid line) and (88) (dashed
curve) for for $\a$ = 1, $\eps$ = .75 and $q$ = 1.

\vs{1cm}

\noindent {\bf Fig. 2} : Plot of imaginary parts of the potentials
in equations (65) (dotted curve), (82) (solid line) and (88)
(dashed curve) for for $\a$ = 1, $\eps$ = .75 and $q$ = 1.

\vs{1cm}

\noindent {\bf Fig. 3} : Plot of real parts of the potentials in
(65) (dotted curve), (98) (solid curve) and (104) (dashed curve)
$\a$ = 1, $\eps$ = .5 and $q$ = 1.

\vs{1cm}

\noindent {\bf Fig. 4} : Plot of imaginary parts of the potentials
in (65) (dotted curve), (98) (solid curve) and (104) (dashed
curve) $\a$ = 1, $\eps$ = .5 and $q$ = 1.

\vs{1cm}

\noindent {\bf Fig. 5} : Plot of real parts of the potentials in
equations (110) (dotted curve), (120) (solid curve) and (132)
(dashed curve) for $\lambda$ = 25, $\mu$ = 5.

\vs{1cm}

\noindent {\bf Fig. 6} : Plot of imaginary parts of the potentials
in equations (110) (dotted curve), (120) (solid curve) and (132)
(dashed curve) for $\lambda$ = 25, $\mu$ = 5. \pb
\begin{figure}[h]
    \centering
        \includegraphics{C:/WINDOWS/Desktop/fig1.eps}
    \caption{.Plot of real parts of the potentials in equations (65) (dotted curve), (82) (solid line) and (88) (dashed curve) for $\a$ = 1, $\eps$ = .75 and $q$ = 1.}
\end{figure}
\begin{figure}[h]
    \centering
        \includegraphics{C:/WINDOWS/Desktop/fig2.eps}
    \caption{Plot of imaginary parts of the potentials in equations (65) (dotted curve), (82) (solid line) and (88) (dashed curve) for $\a$ = 1, $\eps$ = .75 and $q$ = 1.}
\end{figure}

\pb
\begin{figure}[h]
    \centering
        \includegraphics{C:/WINDOWS/Desktop/fig3.eps}
    \caption{Plot of real parts of the potentials in (65) (dotted curve), (98) (solid curve) and (104) (dashed curve) for $\a$ = 1, $\eps$ = .5 and $q$ = 1.}
\end{figure}
\begin{figure}
    \centering
        \includegraphics{C:/WINDOWS/Desktop/fig4.eps}
    \caption{Plot of imaginary parts of the potentials in (65) (dotted curve), (98) (solid curve) and (104) (dashed curve) for $\a$ = 1, $\eps$ = .5 and $q$ = 1.}
\end{figure}

\pb
\begin{figure}[h]
    \centering
        \includegraphics{C:/WINDOWS/Desktop/fig5.eps}
    \caption{Plot of real parts of the potentials in equations (110) (dotted curve), (120) (solid curve) and (132) (dashed curve) for $\lambda$ = 25, $\mu$ = 5.}
\end{figure}
\begin{figure}[h]
    \centering
        \includegraphics{C:/WINDOWS/Desktop/fig6.eps}
    \caption{Plot of imaginary parts of the potentials in equations (110) (dotted curve), (120) (solid curve) and (132) (dashed curve) for $\lambda$ = 25, $\mu$ = 5.}
\end{figure}